\documentstyle[epsf]{article}

 \renewcommand{\theequation}{\thesection.\arabic{equation}}

 \newcommand{\bm}[1]{\mbox{\boldmath $#1$}}
 \newcommand{\TT}{{\bm T}}
 \newcommand{\KK}{{\bm K}}
 
 \newcommand{\AAA}{{\bm A}}
 \newcommand{\BB}{{\bm B}}
 \newcommand{\MM}{{\bm M}}
 \newcommand{\RRR}{{\bm R}}
 \newcommand{\UU}{{\bm U}}
 \newcommand{\sig}{{\bm \sigma}}
 \newcommand{\vsig}{{\vec{\bm \sigma}}}
 \newcommand{\nn}{{\vec{n}}}
 \newcommand{\zr}[1]{\mbox{\hspace*{#1em}}}
 \newcommand{\ID}{\mbox{{\sf 1}\zr{-0.16}\rule{0.04em}{1.55ex}\zr{0.1}}}
 \newcommand{\NN}{\mbox{\zr{0.1}\rule{0.04em}{1.6ex}\zr{-0.05}{\sf N}}}
 \newcommand{\RR}{\mbox{\zr{0.1}\rule{0.04em}{1.6ex}\zr{-0.05}{\sf R}}}
 \newcommand{\CC}{\mbox{\zr{0.1}\rule{0.04em}{1.6ex}\zr{-0.30}{\sf C}}}
 \newcommand{\ZZ}{\mbox{\sf Z\zr{-0.45}Z}}
 \newcommand{\which}{\mbox{\hspace{2mm}}|\mbox{\hspace{2mm}}}
 \newcommand{\tr}{\mbox{tr}}
 \newcommand{\rultb}{\rule[-8mm]{0mm}{18mm}}

\begin{document}
 \bibliographystyle{unsrt}

 \parindent 0pt
 
 \begin{center}
 {\Large {\bf
 TRACE MAPS, INVARIANTS, AND \\
 \vspace{2mm}
 SOME OF THEIR APPLICATIONS}}
 \end{center}
 \vspace{5mm}
 
 \begin{center}
 {\large\sc M. Baake$^{1)}$, U. Grimm$^{2)}$, and D. Joseph$^{1)}$}

 \vspace{5mm}

 {\small
 1) Institut f\"ur Theoretische Physik, Universit\"at T\"ubingen, Auf der
 Morgenstelle 14, \linebreak  D-7400 T\"ubingen, Germany
 
 2) Department of Mathematics, The University of Melbourne, Parkville
 VIC 3052, Australia}
 
 \vspace{5mm}
 \end{center}
 \vspace{5mm}
 
\begin{quote}
Trace maps of two-letter substitution rules are investigated with special
emphasis on the underlying algebraic structure and on the existence of
invariants. We illustrate the results with the generalized Fibonacci chains
and show that the well-known Fricke character
\mbox{$I(x,y,z) = x^2+y^2+z^2-2xyz-1$} is not the only type of invariant that can
occur. We discuss several physical applications to electronic spectra
including the gap-labeling theorem, to kicked two-level systems, and to the
classical 1D Ising model with non-commuting transfer matrices.
\end{quote} 
 

\parindent15pt
\vspace{5mm}

\section{Introduction}
\setcounter{equation}{0}
\label{sec1}

Trace maps have been introduced to calculate the spectrum of 1D Kronig-Penney
and related models on non-periodic structures like the Fibonacci chain
\cite{KKT,Siggia}. In the sequel, these maps have found a broad variety of
applications, ranging from electronic and phononic structure to 1D scattering
and transport (cf.\ \cite{Kohmoto,Iochum,BJK,Bellissard} and references
therein),
from a spin in a magnetic field to kicked two-level systems \cite{Luck1,Graham},
and from classical to quantum spin systems \cite{Luck,Sutherland,You}.
Within only a couple of years, the literature on these subjects thus became
widespread and hard to survey.

On the other hand, there is a considerable amount of activity on the
mathematical aspects of trace maps, ranging from the systematic treatment
of substitution rules \cite{Allouche}, through computer algebraic
investigations \cite{Ali} to powerful analytic
and algebraic results \cite{Peyriere,Iguchi,Belli1,Kramer,Frank,Sire91}, 
and to the
treatment of trace maps as dynamical systems as such \cite{Holzer,Roberts}.
Unfortunately, several if not many of the exact results have not found their
way to the explicit treatment of physical systems, partially because they
escaped notice and partially because they require some non-trivial
mathematics like, e.g., the general gap labeling theorem for
Schr\"odinger operators \cite{Belli1,Simon}.

It is this gap between the theory of trace maps and their application which we
would like to bridge, and we hope that the present article might prove useful
for that. It is important to find a systematic, but simple formulation of the
trace maps. By this we mean a formulation that is close to that of the
Fibonacci case (being the best known) but brings no artificial complications
for the generalizations (like singularities induced by coordinate systems).
It turns out that the formulation of \cite{Allouche} is very appropriate.
Here, all two-letter substitution rules (and we will only consider those)
lead to trace maps that are mappings
$F$ from $\CC^3$ to $\CC^3$ where the component mappings are polynomials with
integral coefficients only. No poles are present as in alternative
formulations with rational functions (cf.\ \cite{You}), and invariants can be
described more easily. Therefore, we will follow this path and
occasionally link it to other formulations.

Let us now briefly outline how the remainder of this article is organized.
In Sec.\ 2, we start with some basic definitions and properties of trace
maps, using the algebraic approach via the free group of two generators
\cite{Peyriere,Kramer}. We then formulate the invariance and transformation
properties of the Fricke character $I(x,y,z) = x^2+y^2+z^2-2xyz-1$
\cite{Fricke}, a proof
of which is given in the Appendix. We illustrate the results with several
examples and discuss, in Sec.\ 3, the generalized Fibonacci chains in some
more detail.

Sec.\ 4 deals with the application to electronic structure, where we apply
the gap labeling theorem to the generalized Fibonacci class. For a subclass
of models, we also present an intuitive, direct approach which does not
rely on the abstract result and, additionally, gives an indexing scheme of
the gaps of the approximants. 
In Sec.\ 5, we very briefly describe the application to a
kicked two-level system which follows a quasiperiodic SU(2) dynamics. Sec.\ 6
then deals with the classical 1D Ising model with coupling constants
and magnetic fields varying according to the non-periodic chain. Here, the
trace map gives rise to a direct iteration of the partition function
which allows at least an effective numerical treatment of the case of 
non-commuting transfer matrices. This is followed by the concluding remarks in
Sec.\ 7 and an Appendix where we prove the general transformation law
of the Fricke character $I(x,y,z)$ as given in Eq.~(\ref{2.16}).

\section{Trace maps and their properties}
\setcounter{equation}{0}
\label{sec2}

To explain what a trace map of a two-letter substitution rule is, and what
its main properties are in simple but general terms, it is the best to start
from the {\em free group} generated by two generators $a, b$, i.e., from
\begin{equation}
\label{2.1}
F_2 := <\!<a,b>\!> .
\end{equation}
Its elements consists of all possible finite words $w = w(a,b,a^{-1},b^{-1})$.
$F_2$ is an infinite group, multiplication is formal composition of words,
and the empty word, $e$, is the neutral element.

A {\em substitution rule} $\varrho$ now is a mapping of $F_2$ into itself,
where - for obvious reasons - we are only interested in homomorphisms,
i.e., in $\varrho$'s with the property
\begin{equation}
\label{2.2}
\varrho(w_1w_2) = \varrho(w_1)\varrho(w_2) .
\end{equation}
Here, the multiplication is the group multiplication in $F_2$. To specify a
homomorphism, it is therefore sufficient to give the images of the two
generators,
\begin{equation}
\label{2.3}
{\varrho} :
\begin{array}{ccc}
          a & \rightarrow & w_a \\
          b & \rightarrow & w_b
\end{array}
\end{equation}
where $w_a$ and $w_b$ are words in $a,b,a^{-1},b^{-1}$, for details we refer
to \cite{Magnus,Kramer}. The concatenation of two homomorphisms,
$\varrho_2 \circ \varrho_1$, is well defined. However, one has to {\em insert}
the rule of $\varrho_1$ into that of $\varrho_2$ which is a bit against the
usual practice. Therefore, we follow \cite{Peyriere} and define the
multiplication
\begin{equation}
\label{2.31}
\varrho_1\varrho_2 = \varrho_1 \cdot \varrho_2 := \varrho_2 \circ \varrho_1 .
\end{equation}
With this multiplication, the set of homomorphisms becomes a monoid with the
identity as neutral element. We call this monoid $\Theta_2 = \mbox{Hom}(F_2)$.

A special role is played by the invertible homomorphisms $\varrho$, i.e.,
by {\em automorphisms}. They form a group $\Phi_2$, that is also finitely
(but not freely) generated \cite{Nielsen1,Neumann,Magnus}
\begin{equation}
\label{2.4}
\Phi_2 := <\!<U,\sigma,P>\!>
\end{equation}
where the generators are defined by

\vspace{2mm}
\begin{center}
\begin{tabular}{|c|c|c|c|}
\hline
{\rule[-3mm]{0mm}{8mm}}
$w$ & $U(w)$ & $\sigma(w)$ & $P(w)$ \\
\hline
\rule[-1mm]{0mm}{6mm}
 $a$ & $ab$ & $a^{-1}$ & $b$ \\
\rule[-3mm]{0mm}{6mm}
 $b$ & $b$  & $b$      & $a$ \\
\hline
\end{tabular}
\end{center}
\vspace{2mm}
For the relations amongst the generators, we refer to p.\ 164 of \cite{Magnus}.

Automorphisms transform the {\em group commutator} $\KK(a,b) := aba^{-1}b^{-1}$
to a conjugate either of itself or of its inverse
\begin{equation}
\label{2.6}
\varrho(\KK(a,b)) = \KK(\varrho(a),\varrho(b)) = w \cdot [\KK(a,b)]^{\pm1}
\cdot w^{-1} ,
\end{equation}
where $w$ is an element of $F_2$ and $(\KK(a,b))^{-1} = \KK(b,a)$. This can
directly be verified by means of the generators, but has also
been shown independently \cite{Nielsen2}. As we will see, this does 
not extend to $\varrho \not\in \Phi_2$.

For the characterization of a substitution rule, it is useful to define
the corresponding {\em substitution matrix}. First, one maps the elements
of $F_2$ to lattice points in the square lattice, $\ZZ^2$, by simply adding
the powers of $a$ and the powers of $b$ separately in the word considered.
This powercounting induces a homomorphism from $\Phi_2 = \mbox{Aut}(F_2)$ to
Gl$(2,\ZZ) = \mbox{Aut}(\ZZ^2)$ \cite{Magnus} and, even more, from the monoid
$\Theta_2 = \mbox{Hom}(F_2)$ to Hom$(\ZZ^2) = \mbox{Mat}(2,\ZZ)$, the latter 
being the monoid of 2x2-matrices with integral elements. Note that
Gl$(2,\ZZ)$ consists of
precisely those matrices $\RRR \in \mbox{Hom}(\ZZ^2)$ with det$(\RRR) = \pm1$,
and it is this determinant which determines the exponent in Eq.~(\ref{2.6}).

We call $\RRR_{\varrho}$ the substitution matrix of the substitution
rule $\varrho$, if we can read the number of $a$'s and $b$'s in the words
$w_a$ and $w_b$ rowwise, i.e.,
\begin{equation}
\label{2.7}
\RRR_{\varrho} =
\left( \begin{array}{ll}
 \#_a(w_a) & \#_b(w_a) \\
 \#_a(w_b) & \#_b(w_b) \\
\end{array} \right)
\end{equation}
Note that often the transpose of $\RRR_{\varrho}$ is used. With our definition
of the product $\varrho\sigma$, see Eq.~(\ref{2.31}), we have indeed
\begin{equation}
\label{2.9}
\RRR_{\varrho\sigma} = \RRR_{\varrho} \cdot \RRR_{\sigma} ,
\end{equation}
and, e.g., the Fibonacci rule reads $a \rightarrow b,\hspace{1mm} b
\rightarrow ba \hspace{1mm}$ and leads to the matrix
\begin{equation}
\label{2.8}
\RRR_{\varrho} =
\left( \begin{array}{ll}
  0 & 1 \\
  1 & 1 \\
\end{array} \right).
\end{equation}

Of course, different substitution rules can result in the same substitution
matrix and the kernel of the homomorphism from $\Theta_2$ to Mat$(2,\ZZ)$
contains all inner automorphisms of $\Phi_2$: if $\varrho$ acts as
$\varrho(a) = w a w^{-1}$ and $\varrho(b) = w b w^{-1}$, one has
$\RRR_{\varrho} = \ID$.
The kernel of the mapping from $\Phi_2$ to Gl$(2,\ZZ)$ consists of precisely
these inner automorphisms \cite{Nielsen1} while there are further elements
in the general case.

So far, we have only set up the algebraic background. To come to the trace
maps, we have to interpret the substitution in the free group as a substitution
--- and thus a recursion --- in the group Sl$(2,\CC)$ of 2x2 complex unimodular
matrices. We simply insert two matrices $\AAA, \BB$ for the letters $a, b$ and
continue. But Sl$(2,\CC)$ matrices have very special properties as a
consequence of the Cayley-Hamilton theorem of linear algebra, namely 
\begin{equation}
\label{2.10}
\begin{array}{rcl}
\AAA^2          & = & \tr(\AAA)\AAA - \ID \\
\AAA + \AAA^{-1} & = & \tr(\AAA)\ID \\
\AAA^n          & = & U_{n-1}(x)\AAA - U_{n-2}(x)\ID
\end{array}
\end{equation}
where $x = \frac{1}{2}\tr(\AAA)$ and $U_n(x)$ are Chebyshev's polynomial
of the second kind \cite{Abramowitz} defined by:
\begin{equation}
\label{2.11}
\begin{array}{ccl}
U_{-1}(x) & \equiv  & 0, \hspace{5mm} U_0(x) \equiv 1 \\
U_{n+1}(x)&  =      & 2xU_n(x) - U_{n-1}(x)  \; . \\
\end{array}
\end{equation}
For later use, we run the recursion also backwards and formally obtain
polynomials $U_n$ for all $n \in \ZZ$. Also, Eq.~(\ref{2.10}) is then valid
for all integers $n$.

We now define
\begin{equation}
\label{2.12}
x = \frac{1}{2}\tr(\AAA), \hspace{5mm}
y = \frac{1}{2}\tr(\BB), \hspace{5mm}
z = \frac{1}{2}\tr(\AAA\BB).
\end{equation}
Then, as a result of Eq.~(\ref{2.10}) and of the general identities
$\tr(\AAA + \BB) = \tr(\AAA) + \tr(\BB)$ and 
$\tr(\AAA\BB) = \tr(\BB\AAA)$, we
find that the traces after one iteration of an arbitrary substitution rule
$\varrho$ can be expressed by $x, y, z$ again \cite{Allouche}. This is the
corresponding trace map
\begin{equation}
\label{2.13}
F_{\varrho} : \hspace{2mm}
\left( \begin{array}{c}
 x \\
 y \\
 z
\end{array} \right)
\hspace{2mm}
\rightarrow
\hspace{2mm}
\left( \begin{array}{c}
 f_{\varrho}(x,y,z) \\
 g_{\varrho}(x,y,z) \\
 h_{\varrho}(x,y,z) \\
\end{array} \right)
\end{equation}
where $f_{\varrho}, g_{\varrho}, h_{\varrho} \in \ZZ[x,y,z]$ are
three polynomials
with integral coefficients. All trace maps defined that way are thus
$C^{\infty}$ mappings of $\CC^3$ into itself - and interesting dynamical
systems as such. Furthermore we find
\begin{equation}
\label{2.14}
F_{\varrho\sigma} = F_{\varrho} \circ F_{\sigma},
\end{equation}
compare also \cite{Peyriere}. The mapping from the substitution rules to the
trace maps is a homomorphism between monoids. For the mapping from $\Phi_2$ to
Diff($\CC^3$), the kernel is known, compare \cite{Roberts}. The mapping
induces also a homomorphism from Gl(2,$\ZZ$) to the trace maps,
the kernel of which is $\{\pm\ID\}$.

One of the most celebrated properties of trace maps is the invariance of the
polynomial
\begin{equation}
\label{2.15}
I(x,y,z) = x^2 + y^2 + z^2 - 2xyz - 1
\end{equation}
under all trace maps that stem from an automorphism
\cite{KKT,Ali,Peyriere,Kramer}.
In our formulation, it follows either immediately from the
Nielsen theorem (Eq.~(\ref{2.6})) (observing that $\AAA \in \mbox{Sl}(2,\CC)$
implies $\tr(\AAA) = \tr(\AAA^{-1})$) or, alternatively, from the
generators $U, \sigma,
P$ in Eq.~(\ref{2.3}) and their trace maps by simple substitution into
Eq.~(\ref{2.15}):

\vspace{2mm}
\begin{center}
\begin{tabular}{|c|c|l|}
\hline
{\rule[-3mm]{0mm}{8mm}}
$\varrho$ & $\RRR_{\varrho}$ & $\hspace{17mm}F_{\varrho}$ \\
\hline
$U$ \rultb   & 
$ \left( \begin{array}{cc}
 1 & 1 \\
 0 & 1 \\
\end{array} \right) $ & 
$ \left( \begin{array}{c}
 x \\
 y \\
 z \\
\end{array} \right)
\hspace{2mm}
\rightarrow
\hspace{2mm}
\left( \begin{array}{c}
 z \\
 y \\
 2yz-x \\
\end{array} \right) $
 \\
\hline
$\sigma$ \rultb &
$ \left( \begin{array}{cc}
 -1 & 0 \\
  0 & 1 \\
\end{array} \right) $
  & 
$\left( \begin{array}{c}
 x \\
 y \\
 z \\
\end{array} \right)
\hspace{2mm}
\rightarrow
\hspace{2mm}
\left( \begin{array}{c}
 x \\
 y \\
 2xy-z \\
\end{array} \right)$
 \\
\hline
$P$ \rultb  &
$ \left( \begin{array}{cc}
 0 & 1 \\
 1 & 0 \\
\end{array} \right) $  & 
$ \left( \begin{array}{c}
 x \\
 y \\
 z \\
\end{array} \right)
\hspace{2mm}
\rightarrow
\hspace{8mm}
\left( \begin{array}{c}
 y \\
 x \\
 z \\ 
\end{array} \right) $
 \\
\hline
\end{tabular}
\end{center}
\vspace{2mm}

This remarkable property appeared in the Fibonacci case in \cite{KKT},
but the theory of Fricke characters, where $I(x,y,z)$ belongs to, is much older
(cf.\ \cite{Kramer} and references therein): It goes back to the last
century \cite{Fricke}.

But this is not the end of the story because one can show that the Fricke
character follows a specific transformation law under {\em any} substitution 
rule, not just under automorphisms. Indeed, for $\varrho \in \mbox{Hom}(F_2)$,
one finds
\begin{equation}
\label{2.16}
I(F_{\varrho}(x,y,z)) = P_{\varrho}(x,y,z) \cdot I(x,y,z)
\end{equation}
where $P_{\varrho}(x,y,z) \in \ZZ[x,y,z]$ is a polynomial with integral
coefficients, called the transformation polynomial of $\varrho$.
This relation was conjectured on the base of computer
algebraic manipulations \cite{Ali} and proved shortly after \cite{Peyriere}.
In the Appendix, we give an independent and slightly more complete proof of it.

This relation, together with the general properties mentioned above,
has a number of strong consequences. The reader is invited
to find some himself. We list the following:

\begin{itemize}
\item The variety ${\cal M}_0 = \{ (x,y,z) \in \CC^3 \which
I(x,y,z) = 0 \}$ is an invariant surface for all trace maps. 
\item $(1,1,1)$ is a fixed point and $\{ (1,-1,-1), (-1,1,-1),
(-1,-1,1) \}$ is an invariant set of all trace maps. 
\item If $(0,0,0)$ is not a fixed point of $F_{\varrho}$, its image lies
on ${\cal M}_0$. 
\item If $\varrho \in \Phi_2$, $(0,0,0)$ is a fixed point of
$F_{\varrho}$. Conversely, $F_{\varrho}(0,0,0) = (0,0,0)$ implies
$P_{\varrho}(0,0,0) = 1$ and $\nabla P_{\varrho}(0,0,0) = (0,0,0)$, 
but not $P_{\varrho} \equiv 1$. 
\item $P_{\varrho}(0,0,0)$ is either $0$ or $1$. 
\item If $P_{\varrho}$ is constant, we have either
$P_{\varrho} \equiv 1$ or $P_{\varrho} \equiv 0$. 
\item If $\varrho$ is of finite order, $\varrho \in \Phi_2$
and $P_{\varrho} \equiv 1$. 
Furthermore, one has $\varrho^{n}=e$ with
$n=1,2,3,4$, or $6$ and $F_{\varrho}^{m}=\mbox{Id}$ with $m=1,2$, or $3$.
\item The substitution $\varrho$ is invertible if and only if
$P_{\varrho} \equiv 1$. 
\item The substitution $\varrho$ has nontrivial kernel if and only if
$P_{\varrho} \equiv 0$. 
\item The substitution $\varrho$ is injective but not onto if and
only if $P_{\varrho} \not\equiv$ const.
\item The set $\{F_{\varrho}\; |\; \varrho\!\in\!\Phi_{2}\}$ of invertible trace maps
is a group isomorphic to PGl(2,$\ZZ$).
\end{itemize}

We cannot give the proofs here, many of which can be found in the work of
Peyri\`ere and coworkers, compare \cite{Peyriere2} and references therein.
Let us only remark that one takes profit
from using trace orbits of SU(2) matrices and some standard arguments from
calculus.

Instead, we would like to remark that, for all trace maps, the dynamics is
completely integrable on ${\cal M}_0$. Any triple $(x,y,z)$ can be obtained
as traces of diagonal matrices, cf.\ Appendix. Those in turn can be written
as diag$(\exp{(i\alpha)},\exp{(-i\alpha)})$, $\alpha \in \CC$, wherefore the
multiplication is easily done by counting the exponents of the matrices
involved. For any approximant, this can be expressed by the eigenvalues of
the substitution matrix $\RRR_{\varrho}$ and is a straightforward generalization
of the corresponding property of the Fibonacci chain.
This way, one sets up an interesting relation to the theory of pseudo
Anosov maps, see \cite{Roberts,MacKay} for details.

\section{The generalized Fibonacci chains
 and new invariants}
\setcounter{equation}{0}
\label{sec3}

In this chapter, we will illustrate some of the properties in the class of
generalized Fibonacci chains. The latter are defined by the substitution
rules
\begin{equation}
\label{3.1}
\varrho^{(k,\ell)} : \hspace{2mm}
\left( \begin{array}{l}
 a \rightarrow b \\
 b \rightarrow b^{\ell}a^k
\end{array} \right) ,
\hspace{5mm}
\RRR_{\varrho}^{(k,\ell)} =
\left( \begin{array}{cc}
 0 & 1 \\
 k & \ell
\end{array} \right)
\end{equation}
where $k, \ell \in\ZZ$.
If the context is clear, we will frequently drop the index $(k,\ell)$.
$\RRR_{\varrho}$ is unimodular if and only if $k = \pm1$. This coincides with
$\varrho$ being an automorphism, i.e., invertible. From
det$(\RRR_{\varrho}) = -k$ we also see that $k \not= 0$ gives non-singular
cases.

What are the corresponding trace maps? A one-page calculation results in
\begin{equation}
\label{3.2}
F_{\varrho} : \hspace{2mm}
\left( \begin{array}{c}
 x \\
 y \\
 z
\end{array} \right)
\hspace{2mm}
\rightarrow
\hspace{2mm}
\left( \begin{array}{c}
 y \\
 g(x,y,z) \\
 h(x,y,z) \\
\end{array} \right)
\end{equation}
with
$g(x,y,z) = U_{k-1}(x)U_{\ell-1}(y) \cdot z - U_{k-1}(x)U_{\ell-2}(y) \cdot x 
- U_{k-2}(x)U_{\ell-1}(y) \cdot y + U_{k-2}(x)U_{\ell-2}(y)$
and $h(x,y,z) = g(x,y,z)|_{\ell \rightarrow \ell+1}$.

This is equivalent to the formulation of \cite{You}. There, the iteration
is given as a difference equation of the form 
$y_{n+1} = r(y_n,y_{n-1},y_{n-2})$ with rational function $r$.
The latter has singularities as a pure consequence of the coordinates chosen.
Therefore, Eq.~(\ref{3.2}) is advantageous.
Another page of calculation establishes the transformation polynomial to be
\begin{equation}
\label{3.3}
P_{\varrho}(x,y,z) = (U_{k-1}(x))^2.
\end{equation}
The cases $k = 1$ and $k = -1$ yield $P_{\varrho} \equiv 1$
(and hence the invariance of $I(x,y,z)$) which we know already from the
invertibility of $\varrho$ in these cases (independently of $\ell$). 
The substitution
rule $\varrho$ has nontrivial kernel for $k = 0$ (e.g., the kernel contains
$a^{-\ell}b$) wherefore $P_{\varrho} \equiv 0$ is the necessary consequence.
The remaining cases ($k \not\in \{-1,0,1\}, \ell \in \ZZ$) belong to the
class of
injective, but non-invertible substitutions. Many properties of Sec.\ 2
can therefore be demonstrated within this class of examples.

One interesting question is the existence of further invariants. That this
indeed is possible can be seen from the case $\ell = 1, k = 2$ where we find
\begin{equation}
\label{3.4}
F_{\varrho} : \hspace{2mm}
\left( \begin{array}{c}
 x \\
 y \\
 z
\end{array} \right)
\hspace{2mm}
\rightarrow
\hspace{2mm}
\left( \begin{array}{c}
 y \\
 2xz - y\\
 4xyz -2x^2 - 2y^2 + 1 \\
\end{array} \right).
\end{equation}
As can be checked directly, this leaves the polynomial $H$ invariant,
\begin{equation}
\label{3.5}
H(x,y,z) = (4x^2 - 1)y - 2xz
\end{equation}
This could be expected from the close relationship of Eq.~(\ref{3.4}) to the
period doubling map \cite{Luck2,Ali,Bovier} where an analogous invariant
exists. $H$ is different from $I$ and does in fact foliate the single
invariant sheet $\{ I(x,y,z) = 0 \}$, see Fig.\ 1.

A closer look shows that the 1D lines on $\{ I = 0 \}$ are an immediate
consequence of $\RRR_{\varrho}$ having an eigenvalue of modulus one.
The motion of
the dynamical system on $\{ I = 0 \}$ is integrable and related to torus maps.
But if there is a direction neither contracted nor expanded, one could expect
another invariant of the trace map $F_{\varrho}$ which foliates at least the
surface $\{ I = 0 \}$ and, eventually, even the whole $\CC^3$.

Now, our substitution matrix $\RRR_{\varrho}$ has eigenvalues
$\lambda_{\pm} = \frac{1}{2} (\ell \pm \sqrt{\ell^2 + 4k})$.
If they are rational numbers, they are automatically integers. This happens
if and only if
\begin{equation}
\label{3.6}
k = m \cdot \ell + m^2
\end{equation}
with $m$ integer, where we find $\lambda_+ = \ell + m$ and $\lambda_- = -m$.
Eq.~(\ref{3.4}) corresponds to $m = 1$, and it turns out that it is not a
singular case. Indeed, for $k = \ell + 1$ in Eq.~(\ref{3.2}), we find the
amazingly simple invariant
\begin{equation}
\label{3.7}
H(x,y,z) = U_{\ell+1}(x) \cdot y - U_{\ell}(x) \cdot z .
\end{equation}
This is a generalization of Eq.~(\ref{3.5}), for a proof one needs the
identities $U_{\ell}^2 - U_{\ell-1}U_{\ell+1} \equiv 1$ and
$U_{\ell}U_{\ell-1} - U_{\ell+1}U_{\ell-2} \equiv U_1$. Also for the case
$m = -(\ell+1)$, Eq.~(\ref{3.7}) gives the right answer, only $\lambda_+$
and $\lambda_-$ are interchanged.

The remaining cases are $m = -1$ and $m = 1 - \ell$. Both result in
$k = 1 - \ell$ and are equivalent. From the extension of the recursion,
Eq.~(\ref{3.2}), to negative indices, we know that 
$U_{-(n+2)} = -U_n$ for $n \in \ZZ$. By means of the identity
$U_{l-1}^2 - U_lU_{l-2} \equiv 1$ one can show the invariance of the polynomial
\begin{equation}
\label{3.8}
\tilde{H}(x,y,z) = U_{\ell-1}(x) \cdot y - U_{\ell-2}(x) \cdot z
= -(U_{-\ell-1}(x) \cdot y - U_{-\ell}(x) \cdot z) .
\end{equation}

This again is a very simple expression and resembles that of Eq.~(\ref{3.7}).
We think that no further invariants exist in the class of models investigated
here but could not finally settle this question. We can combine the two
examples shown formulating $k=1\pm\ell$ and
$H^{(\pm)}(x,y,z) = \pm(U_{\pm(\ell+1)}(x)\cdot y - U_{\pm \ell}(x)
\cdot z)$. This covers all cases in the class of generalized Fibonacci
chains where the corresponding torus map has a unimodular eigenvalue. 

Let us close this section with a short remark why invariants are
$\mbox{important}$.
The mappings are three dimensional and could show the dynamics of 3D
discrete dynamical systems. If an invariant exists, they locally look like
2D systems and inherit the well understood properties of them like period
doubling routes to chaos etc., compare the Appendix of \cite{Roberts}.

Now, the Fricke character is an invariant precisely for the trace maps of
automorphisms, but many interesting trace maps do not stem from automorphisms.
Nevertheless, period doubling can occur (as in the so-called period doubling
map). The existence of invariants like $H$ or $\tilde{H}$ can explain why ---
and a further investigation is in progress.

\section{Electronic spectra and gap labeling}
\setcounter{equation}{0}
\label{sec4}

Amongst the applications of trace maps to physical systems, that to
electronic spectra is perhaps the most widespread one \cite{KKT,Fujiwara}.
The transfer matrices attached to the different intervals describe the
propagation of the amplitudes of the wave function through the chain.
In the case of the continuous Schr\"odinger equation, the transfer matrices
for plane waves belong to SU(1,1) \cite{BJK,Wurtz,Sueto}.
The corresponding trace map acts in $\RR^3$ and allows the description
of the spectrum, while the underlying matrix system can be used to identify
the states. The best way to proceed is to tackle the
full Schr\"odinger equation
\begin{equation}
\label{4.1}
(-\frac{\hbar^2}{2m}\Delta + V(x))\Psi(x) = E\Psi(x)
\end{equation}
where $V(x)$ is the potential on the chain. But things are a little bit
complicated because several quantities become energy dependent. Especially the
very invariant $I(x,y,z)$ shows a significant dependence on the energy of
the incident wave. Therefore, most authors restrict themselves to the
tight-binding case, where Eq.~(\ref{4.1}) is replaced by the discrete version
(in suitable units)
\begin{equation}
\label{4.2}
({\bm {\cal H}}\Psi)_n = \Psi_{n+1} + \Psi_{n-1} + V_n\Psi_n = E\Psi_n,
\end{equation}
where $I(x,y,z) = \frac{1}{4}(V_2-V_1)^2$. 
$V_1, V_2$ are the potentials on the two different intervals.
Although one misses some typical features of the continuous case \cite{BJK},
many generic properties of the spectra remain the same, e.g., in both cases
many spectra are Cantor sets with zero Lebesgue measure (see
\cite{Belli1} for a criterion). The price one has to pay for going
from the full Schr\"odinger equation to the tight-binding approximation is
that in the latter case one needs a whole one parameter set of Hamiltonians
which correspond to the single Hamiltonian in the Schr\"odinger equation
\cite{Belli4}.

Nevertheless, looking at the gap labeling of the spectra, we can make the
restriction to the tight-binding case for simplicity without loss of
information. We emphasize that all results can be extended to
the continuous equation, compare \cite{Belli4}. For the tight-binding
Hamiltonian of Eq.~(\ref{4.2}), the transfer matrices read:
\begin{equation}
\label{4.3}
\TT_i = 
\left( \begin{array}{cr}
 E-V_i & -1 \\
  1    &  0
\end{array} \right),
\end{equation}
and, therefore,
\begin{equation}
\label{4.4}
\begin{array}{cclcl}
 x & = & \frac{1}{2}\tr(\TT_1)           & = & \frac{1}{2}(E-V_1) \\
   &   &                                       &   &                    \\
 y & = & \frac{1}{2}\tr(\TT_2)           & = & \frac{1}{2}(E-V_2) \\
   &   &                                       &   &                    \\
 z & = & \frac{1}{2}\tr(\TT_2\cdot\TT_1) & = & 
\frac{1}{2}(E-V_1)(E-V_2)-1 .
\end{array}
\end{equation}
One could now think of a specific substitution rule $\varrho$
and investigate the spectra by iterative techniques.

Here, we want to follow a slighty different path. First we want to present the
famous gap labeling theorem formulated by Bellissard and coworkers. Because
it uses non-trivial mathematics we will interpret it in simple words and
apply the more concrete version to the well-known case of the
generalized Fibonacci sequences. Afterwards, we will give an intuitive way
of understanding how the gap labeling theorem works.

Bellissard and coworkers proved the following theorem using techniques
of $C^*$ algebras and K-Theory \cite{Belli1}:

Let ${\bm {\cal H}}$ be a Hamiltonian as in (\ref{4.2}) with a 
bounded potential
and ${\cal A}_{\bm {\cal H}}$ the $C^*$ algebra generated by the 
family of operators
obtained from ${\bm {\cal H}}$ by translation. 
The possible values of the Integrated Density of States  (IDOS) on the
gaps of the spectrum are given by
the image of the $K_0$ group of the $C^*$ algebra ${\cal A}_{\bm {\cal H}}$ of the
Hamiltonian ${\bm {\cal H}}$ under the trace per unit volume, $\tau_*(K_0({\cal A}_{\bm {\cal H}}))$.

This is a general result no matter what the dimension $D$ is.
Thinking of the 1D case (\ref{4.2}) with a potential that takes on
only finitely many different values, the gap labeling 
theorem tells us that the possible
values of the IDOS on the gaps are given by all possible frequencies
of all possible words in the infinite chain of the substitution $\varrho$.
But one can express all these
frequencies of words with length greater than one by the 
frequencies of the words of length one and two.
This leads to the "concrete" gap labeling theorem \cite{Belli1}:

Let ${\bm {\cal H}}$ be a Hamiltonian as in (\ref{4.2}) with the potential given
by a primitive substitution, $\varrho$, on a finite alphabet, $A$.
The possible values of the IDOS on the gaps in the spectrum are given by
the $\ZZ(\lambda^{-1})$ module (resp.\ the part contained in [0,1]),
constructed 
by the components of the normalized eigenvectors $v^{(1)}$ and $v^{(2)}$ to the
maximal common eigenvalue $\lambda$ of the substitution matrix for one-letter
words $\MM^{(1)}_{\varrho}$ and for two-letter words $\MM^{(2)}_{\varrho}$.

The substitution $\varrho$ is called irreducible if, for any pair
of letters $a,b$ from the alphabet $A$, the word $\varrho^{k}(a)$
contains $b$ for some $k$. If $k$ can be chosen independently of
the letters $a,b$, then $\varrho$ is called {\em primitive}. 
To guarantee the existence of a (half-)infinite word as a fixed point of
$\varrho$, one considers a substitution which generates an infinite chain 
from every letter of the alphabet $A$. There must be at least one letter, 
$b \in A$ say, so that $\varrho(b)$ begins with $b$, and this letter $b$ must 
appear in every possible chain.

Let us now apply this theorem. For the calculation of $\MM^{(1)}_{\varrho}$
and $\MM^{(2)}_{\varrho}$, we need the substitution of all two-letter words.
Let $\varrho(w) = a_0a_1a_2...a_n$ be a substitution of a word $w$, then
(with the usual notation, cf.\ \cite{Belli1})
\begin{equation}
\label{4.4a}
\varrho_N(w) = (a_0a_1...a_{N-1})(a_1a_2...a_N)...(a_{m-1}a_m...a_{m+N-2})
\end{equation}
is the substitution of an $N$-letter word ($m$ being the total length of
$\varrho(a_0)$ obtained by the powercounting described
in Sec.\ 2). In particular, the two-letter substitution is:
\mbox{$\varrho_2(w) = (a_0a_1)(a_1a_2)...(a_{m-1}a_m)$}.
Our example is the recursion
\begin{equation}
\label{4.5}
\begin{array}{ccl}
 a & \rightarrow & b \\
 b & \rightarrow & b^{\ell}a^k \\
\end{array},
\end{equation}
which leads to the matrix sytem:
$\TT_{n+1} = \TT_{n-1}^k \cdot \TT_n^{\ell}$, compare \cite{BJK,You}.
The matrix $\MM_{\varrho}^{(1)}$ is nothing but the transpose of the
substitution matrix $\RRR_{\varrho}$, i.e.,
\begin{equation}
\label{4.7}
\MM_{\varrho}^{(1)} =  
\left( \begin{array}{cc}
 0 & k \\
 1 & \ell
\end{array} \right),
\hspace{5mm}
\begin{array}{l}
 m_a = 1    \\
 m_b = k+\ell
\end{array} 
\end{equation}
with eigenvalues
\begin{equation}
\label{4.8}
\lambda_{\pm} = \frac{1}{2} (\ell \pm \sqrt{\ell^2 + 4k}).
\end{equation}
Let us introduce $D = k(k+\ell-1)$ and exclude the case $D=0$ ($\varrho$
is reducible for $k=0$ and does not increase the length of any word
for $k+\ell=1$). Then, we can write the normalized 
eigenvector to $\lambda_+$ as
\begin{equation}
\label{4.9}
v^{(1)} = \frac{1}{D}
\left( \begin{array}{ccc}
k(k+\ell) & - & k\lambda_+ \\
 -k       & + & k\lambda_+ 
\end{array} \right).
\end{equation}
Here, normalization is a statistical one, i.e., $\sum_i v_i = 1$.
Also, $\MM_{\varrho}^{(2)}$ is easy to calculate. We find
\begin{equation}
\label{4.10}
\begin{array}{ccc}
 aa & \rightarrow & bb            \\
 ab & \rightarrow & b^{\ell+1}a^k    \\
 ba & \rightarrow & b^{\ell}a^kb       \\
 bb & \rightarrow & b^{\ell}a^kb^{\ell}a^k
\end{array}
\hspace{5mm}
\Rightarrow
\hspace{5mm}
\begin{array}{ccl}
 \varrho(aa) & \rightarrow & (bb)                            \\
 \varrho(ab) & \rightarrow & (bb)                            \\
 \varrho(ba) & \rightarrow & (bb)^{\ell-1}(ba)(aa)^{k-1}(ab) \\
 \varrho(bb) & \rightarrow & (bb)^{\ell-1}(ba)(aa)^{k-1}(ab)
\end{array}
\end{equation}
which means
\begin{equation}
\label{4.11}
\MM_{\varrho}^{(2)} =  
\left( \begin{array}{cccc}
 0 &  0 & k-1    & k-1   \\
 0 &  0 &  1     &  1    \\
 0 &  0 &  1     &  1    \\
 1 &  1 & \ell-1 & \ell-1
\end{array} \right) .
\end{equation}
The eigenvalues are
\begin{equation}
\label{4.12}
\lambda \in \{ 0, 0, \frac{1}{2} (\ell \pm \sqrt{\ell^2 + 4k}) \} .
\end{equation}
The (statistically) normalized eigenvector to $\lambda_+$ reads
\begin{equation}
\label{4.13}
v^{(2)} = \frac{1}{D}
\left( \begin{array}{ccc}
 (k+\ell)(k-1) & - & (k-1)\lambda_+ \\
 (k+\ell)      & - & \lambda_+     \\
 (k+\ell)      & - & \lambda_+     \\
-(2k+\ell)     & + & (k+1)\lambda_+
\end{array} \right),
\end{equation}
because of
$1/\lambda_+ = (\lambda_+-\ell)/k$, and $\lambda_+^2 = k+\ell\lambda_+$.

Now the calculation of the frequency module has to be done. This requires some
care. First, we observe that the components of $v^{(1)}$ can be obtained by
integral linear combinations of components of $v^{(2)}$, and the latter are
all of the form 
\mbox{$\frac{1}{D}[(m((k+\ell)-\lambda_+) + n(-(2k+\ell)+(k+1)\lambda_+)], 
\hspace{1mm} m,n \in \ZZ$}. This can be rewritten as
\begin{equation}
\label{4.13a}
\frac{1}{D}(\tilde{\mu} + \tilde{\nu}\lambda_+), \hspace{5mm} \tilde{\mu},
\tilde{\nu} \in \ZZ
\end{equation}
but now with the additional constraints
\begin{equation}
\label{4.13b}
\begin{array}{ccc}
 \tilde{\mu} + \tilde{\nu}         & \equiv & 0 \hspace{1mm} (D) \\
 \tilde{\mu} + (k+\ell)\tilde{\nu} & \equiv & 0 \hspace{1mm} (D)
\end{array}.
\end{equation}
This is necessary to guarantee $m,n \in \ZZ$ in the preceeding expression.
Remember that $1/\lambda_+ = (\lambda_+-\ell)/k$.
It is easy to check that multiplication with
$(\lambda_+-\ell)$ leads to new numbers $\tilde{\mu}', \tilde{\nu}'$ which also
fulfil Eq.~(\ref{4.13b}). Consequently, the $\ZZ(\lambda_+^{-1})$ module turns
out to be
\begin{equation}
\label{4.14}
\tau_*(K_0({\cal A}_{\bm {\cal H}})) = \{ \frac{1}{D}\frac{\tilde{\mu} + 
\tilde{\nu}\lambda_+}{k^p} \which
\tilde{\mu},\tilde{\nu},p \in \ZZ, \;\;
\begin{array}{ccc}
 \tilde{\mu} + \tilde{\nu}         & \equiv & 0 \hspace{1mm} (D) \\
 \tilde{\mu} + (k+\ell)\tilde{\nu} & \equiv & 0 \hspace{1mm} (D)
\end{array} \}.
\end{equation}
for the continuous Schr\"odinger equation and $\tau_*(K_0({\cal A}_{\bm {\cal H}})) \cap
[0,1]$ for the tight-binding case.

One has to keep in mind that these are only all {\em possible} values the IDOS
can take. The gap labeling theorem does not tell us whether all gaps are
really open. The most prominent example where gaps are closed systematically
is the Thue-Morse sequence \cite{Luck2,Belli5}. But there, the eigenvalues 
$\lambda_{\pm}$ are
integers, and one has to check whether the frequency module really requires
all linear combinations of the components of $v^{(1)}$ and $v^{(2)}$ or not.
The analogous question arises for the generalized Fibonacci chains with
Eq.~(\ref{3.6}) as condition, but we cannot go into details here.

The special subclass of the metallic Fibonacci sequences
$(k=1, \ell \in \NN)$ has the module 
\begin{equation}
\label{4.14a}
\{ \frac{1}{\ell}(\tilde{\mu} + \tilde{\nu}\lambda_+) \which 
\tilde{\mu},\tilde{\nu} \in \ZZ,
\hspace{5mm} \tilde{\mu}+\tilde{\nu} \equiv 0 \hspace{1mm} (\ell) \} ,
\end{equation} 
whereas the largest eigenvalues
of the substitution matrices are given by the metallic means:
\begin{equation}
\label{4.15}
\lambda_+ = \frac{1}{2}(\ell + \sqrt{\ell^2+4}) = [\ell;\ell,\ell,\ell,...].
\end{equation}
Looking at these chains one is able to give
a more heuristic but intuitive way of understanding how the gap labeling
theorem works.

Let us introduce the generalized Fibonacci numbers, $f_n^{(\ell)}$, by
\begin{equation}
\label{4.16}
f_{n+1}^{(\ell)} = \ell \cdot f_n^{(\ell)} + f_{n-1}^{(\ell)},
\hspace{5mm} f_0^{(\ell)} = 0, \hspace{3mm} f_1^{(\ell)} = 1 .
\end{equation}
This choice of initial conditions will prove useful shortly. We observe that
two consecutive numbers are coprime because the greatest common divisor obeys
gcd$(f_n^{(\ell)},f_{n+1}^{(\ell)}) 
= \mbox{gcd}(f_n^{(\ell)},\ell \cdot f_n^{(\ell)}+f_{n-1}^{(\ell)})
= \mbox{gcd}(f_n^{(\ell)},f_{n-1}^{(\ell)})$,
i.e., coprimality is inherited from the initial conditions.

If we now consider the series of periodic approximants starting from $b$,
i.e., \mbox{$b \rightarrow b^{\ell}a \rightarrow (b^{\ell}a)^{\ell}b$} etc., we know
that these approximants are optimal in the following sense: a given
approximant of length $g_n^{(\ell)}$, where 
\begin{equation}
\label{4.17}
g_n^{(\ell)} = f_n^{(\ell)} + f_{n-1}^{(\ell)},
\end{equation}
is closer to the limit structure than any given other approximant of length
$L \leq g_n^{(\ell)}$. One therefore expects that the structure of the IDOS
of this approximant is as close to the limit IDOS as possible with 
approximants up to that length.
In particular, the values of the IDOS on the gaps converge rapidly.

It is tempting to select series of IDOS plateaus in the sequence of
approximants in such a way that the corresponding gaps belong to each other
(by quantum numbers, strength, symmetry, etc.). Then, if one can find a
labeling, the limit $n \rightarrow \infty$ should reproduce the result of
the general theorem described above. The chains with the metallic means
facilitate this procedure. Here, gaps in consecutive approximants can be
attached to one another by the following simple rule. Given a gap in the
$n$-th approximant. Then, in the step from $n$ to $n+1$, one always grabs the
closest IDOS value possible. If this is not unique (which happens
only occasionally) one takes both possibilities as two branches. One branch
will actually correspond to a new series of gaps, but that is hard to see
in general and does not matter for the limit.

Now comes the trick: the gaps in the $n$-th approximant lead to IDOS values of
the form
\begin{equation}
\label{4.18}
\frac{m}{f_{n}^{(\ell)}+ f_{n-1}^{(\ell)}},
\hspace{5mm} 0 \leq m \leq f_n^{(\ell)} +f_{n-1}^{(\ell)},
\end{equation}
which is an exact result from Bloch theory for the periodic approximants.
But, $m$ can be written as 
\begin{equation}
\label{4.19}
m = \mu \cdot f_n^{(\ell)} + \nu \cdot f_{n-1}^{(\ell)},
\hspace{5mm} \mu, \nu \in \ZZ,
\end{equation}
because $f_n^{(\ell)}$ and $f_{n-1}^{(\ell)}$ are coprime. Taking
$(\mu,\nu)$ as label, it turns
out that Eq.~(\ref{4.19})  selects series of IDOS plateaus --- and hence gaps ---
that follow the rule mentioned before! Furthermore, it is easy to see that
neither possible gaps are missed nor index pairs are missing (as long as
$\mu \cdot f_{n}^{(\ell)} + \nu \cdot f_{n-1}^{(\ell)} \in 
[0,f_n^{(\ell)}+f_{n-1}^{(\ell)}]$). It now is
straightforward to calculate the limit points which gives the set
\begin{equation}
\label{4.20}
\{ \frac{\lambda_+-1}{\ell}(\mu + \nu \frac{1}{\lambda_+}) \which 
\mu, \nu \in \ZZ \} \cap [0,1] .
\end{equation}
A similar argument was recently given for the original Fibonacci chain
\cite{Liu} where this procedure is most simple.
A simple substitution shows that Eq.~(\ref{4.20})
{\em precisely} reproduces the result
obtained above in Eq.~(\ref{4.14a}), including the modulo condition. The labels
obtained from Eq.~(\ref{4.20}) are slightly handier as can be seen in Fig.\ 2,
where, as an example, we show the IDOS of the first approximants of the
'octonacci' chain, ($k = 1, \ell = 2$), and the index pairs $(\mu,\nu)$
starting with the approximants of length 3 and 7.

Let us finally remark that the coincidence of the abstract and the concrete
approach has a strong consequence: As far as we calculated, all gaps of
the approximants involved here are open (for suitable choices of potentials).
But from the second approach, it is meaningful to conjecture that all gaps
are open for the class of metallic Fibonacci chains. For a proof, one could
try pertubative arguments along the lines of \cite{Luck2} because these
chains are quasiperiodic with embedding dimension 2, compare \cite{BJK}.
This will be given elsewhere.

\section{Some properties of kicked 
two-level systems}
\setcounter{equation}{0}
\label{sec5}

As already mentioned in the Introduction, matrices and trace maps derived
from two-letter substitution rules find applications in a variety of problems.
Here, we briefly describe the case of SU(2) dynamics. The latter appears
naturally in the solution of Schr\"odinger's equation for an electron or
another particle with spin $s = 1/2$ in a magnetic field $\vec{B}$ \cite{Haken}.
\begin{equation}
\label{5.1}
i \hbar \dot{\Psi} = \mu_B \vec{B} \vsig \Psi
\end{equation}
with Bohr's magneton $\mu_B = \frac{\hbar}{2m_0}$, Pauli's spin matrices
$\sig_i$, a spinor $\Psi$ of two components, and
\begin{equation}
\label{5.2}
\vec{B} \vsig =
\left( \begin{array}{cc}
 B_3                & B_1 - i B_2 \\
 B_1 + i B_2  & -B_3
\end{array} \right).
\end{equation}
We have written the time-dependent equation in Eq.~(\ref{5.1}) because the
stationary case with constant magnetic field
is completely integrable \cite{Haken} and hence less
interesting than the truly time-dependent one. Here we are interested in
a special class of time-dependent fields that allow the application of
the recursive transfer matrix techniques, where transfer is now in {\em time}.

Rewriting Eq.~(\ref{5.1}) slightly, we obtain the structural core as
\begin{equation}
\label{5.3}
\dot{\Psi} = -i {\bm {\cal H}}(t) \Psi
\end{equation}
with hermitian and traceless operator ${\bm {\cal H}}(t)$. 
The (formal) solution is
\begin{equation}
\label{5.4}
\Psi(t) = \UU(t,t_0) \Psi(t_0)
\end{equation}
with
\begin{equation}
\label{5.5}
\UU(t,t_0) = \mbox{T}[\exp{(-i \int_{t_0}^t {\bm {\cal H}}(\tau)d\tau)}].
\end{equation}
The time evolution operator $\UU$ is a time-dependent SU(2) matrix, and T
denotes time-odering \cite{Field}.

Let us now think of a time sequence following the generalized Fibonacci chain,
cf.~(\ref{4.5}). With $\UU_n := \UU(t_n,t_0)$ we obtain
\begin{equation}
\label{5.6}
\UU_{n+1} = \UU_{n-1}^k \cdot \UU_n^{\ell}
\end{equation}
for the SU(2) dynamics. The treatment of this matrix system gives full
information but goes beyond the scope of the present article. For a summary
of what can
be done with that we refer to \cite{Kramer}. Here, we follow
\cite{Sutherland} and consider the trace systems only which gives insight
into the behaviour of certain observables. Let us formulate it
for the Fibonacci case ($k=\ell=1$), the extension to the other
examples is immediate.

To be explicit in the formulas, we will consider $\delta$-kicks on the time 
intervals, more precisely, at the beginning of them. Any SU(2) matix
$\AAA$ can be written in the form
\begin{equation}
\label{5.7}
\AAA = \exp{(-i a \nn\vsig)} = \cos{(a)} \ID - i \sin{(a)}
\nn\vsig
\end{equation}
with $a \in \RR^+$ and $\nn$ a unit vector (hence, $(\nn\vsig)^2 = \ID$).
From the relation
\begin{equation}
\label{5.8}
[ \nn\vsig , \nn'\vsig ] = 2i \epsilon_{k \ell m} n_k n_{\ell}' \sig_m
\end{equation}
we can conclude that transfer matrices of the elementary intervals certainly
commute if the kicks are in the same direction. But then, the real traces run
on $\{ I = 0 \}$ and stay bounded, $|x| \leq 1, |y| \leq 1, |z| \leq 1$. The
orbit belongs to the pseudo Anosov system, compare \cite{MacKay,Roberts}.
The question now is what happens off this surface.

Consider the kicks $\delta(t) a_j \nn_j\vsig, \hspace{1mm} j = 0,1$ for the
two elementary time intervals. We find
\begin{equation}
\label{5.9}
\UU^{(j)} = \exp{(-i \int_0^{t_j}\delta(t)a_j\nn_j\vsig dt)}
= \cos{(a_j)} \ID - i \sin{(a_j)} \nn_j\vsig
\end{equation}
with
\begin{equation}
\label{5.10}
(\nn_0\vsig) \cdot (\nn_1\vsig) = (\nn_0\nn_1) \ID 
+ i (\nn_0 \times \nn_1) \vsig .
\end{equation}
We can work out the product $\UU^{(2)} = \UU^{(0)}\UU^{(1)}$ and obtain
\begin{eqnarray}
\label{5.11}
\UU^{(2)} & = & \{\cos{(a_0)}\cos{(a_1)} - \sin{(a_0)}\sin{(a_1)} 
\nn_0\nn_1\} \ID \\
& &  + i \{\nn_0\times\nn_1 \sin{(a_0)}\sin{(a_1)}
           - \nn_0 \sin{(a_0)}\cos{(a_1)} - \nn_1 \cos{(a_0)}\sin{(a_1)}\} \vsig \nonumber
\end{eqnarray}
This gives the initial conditions for the traces as
\begin{equation}
\label{5.12}
\begin{array}{l}
x_0 = \cos{(a_0)}, \hspace{4mm} x_1 = \cos{(a_1)},   \\
x_2 = \cos{(a_0)}\cos{(a_1)} - \sin{(a_0)}\sin{(a_1)} (\nn_0\nn_1) .
\end{array}
\end{equation}
With some further calculations one can establish the formula
\begin{equation}
\label{5.13}
I(x_0,x_1,x_2) = [(\nn_0\nn_1)^2 - 1]\cdot(\sin{(a_0)}\sin{(a_1)})^2
\end{equation}
which shows $-1 \leq I \leq 0$, as it must be for SU(2).

One can now start from $I = -1$ and go to $I = 0$ by an adiabatic change of
parameters. This way, one could try to follow (for the
Fibonacci case, say,) the period doubling route to chaos, described recently
in \cite{Roberts}. It is an interesting question whether one could
experimentally find the {\em conservative} feigenvalues predicted theoretically.

\section{Classical 1D Ising model with
non-commuting transfer matrices}
\setcounter{equation}{0}
\label{sec6}

Let us consider a linear chain of $N$ Ising spins
\mbox{$\sigma_{j}\in\{\pm 1\}$},
$j=1,\ldots ,N$, with the energy of a configuration
$\sigma = (\sigma_{1},\sigma_{2},\ldots ,\sigma_{N})$ 
being given by \cite{Baxter} 
\begin{equation}
E(\sigma)\;\; = \;\; 
- \sum_{j=1}^{N} \left( J_{j,j+1} \sigma_{j} \sigma_{j+1}\: +\:
H_{j} \sigma_{j} \right) ,
\label{6.1}
\end{equation}
where we choose periodic boundary conditions, i.e., 
$\sigma_{N+1} = \sigma_{1}$.
The canonical partition function $Z^{(N)}$ of this system is obtained as a 
sum over all possible configurations $\sigma$ as follows
\begin{equation}
Z^{(N)}\;\; = \;\; 
\sum_{\sigma} \prod_{j=1}^{N} \exp \left( K_{j,j+1} 
\sigma_{j}\sigma_{j+1} + h_{j} \sigma_{j} \right) 
\label{6.2}
\end{equation}
with $K_{j,j+1} = J_{j,j+1}/k_{B} T$ and $h_{j} = H_{j}/k_{B} T$.
Here, $T$ is the temperature and $k_{B}$ denotes Boltzmann's constant.
The corresponding free energy per site $F^{(N)}$ is given by
\mbox{$F^{(N)} = - \frac{1}{N} \log Z^{(N)}$} where we absorbed the
factor $1/(k_{B}T)$ into the definition in order to obtain a 
dimensionless quantity.

Eq.~(\ref{6.2}) can be rewritten in the following way
(see \cite{Baxter})
\begin{eqnarray}
Z^{(N)} & = & \sum_{\sigma}\;
{\bm {\cal T}}^{(1,2)}(\sigma_{1},\sigma_{2})\:
{\bm {\cal T}}^{(2,3)}(\sigma_{2},\sigma_{3})\:
\cdots
{\bm {\cal T}}^{(N,1)}(\sigma_{N},\sigma_{1}) \nonumber \\
& = & \tr\left( {\bm {\cal T}}^{(1,2)} {\bm {\cal T}}^{(2,3)} \cdots 
{\bm {\cal T}}^{(N,1)}  \right) \label{6.3} \\
& = & \tr\left( {\bm {\cal T}}^{(N)} \right) \nonumber
\end{eqnarray}
where the elementary transfer matrix ${\bm {\cal T}}^{(j,j+1)}$ is given by the
$2\!\times\! 2$ matrix \cite{Baxter}
\begin{equation}
{\bm {\cal T}}^{(j,j+1)} \;\; = \;\; \left(
\begin{array}{ll} 
\exp\left( K_{j,j+1} + \frac{h_{j} + h_{j+1}}{2} \right) &
\exp \left( - K_{j,j+1} \right) \\
\exp \left( - K_{j,j+1} \right) &
\exp\left( K_{j,j+1} - \frac{h_{j} + h_{j+1}}{2} \right)
\end{array} \right) \;\; ,
\label{6.4}
\end{equation}
while ${\bm {\cal T}}^{(N)}$ denotes the transfer matrix of $N$ sites.
Note that the transfer matrices at different sites in general
do not commute and therefore cannot be diagonalized simultaneously.

So far we have not specified the coupling constants $K_{j,j+1}$
and the magnetic fields $h_{j}$.
A particularly interesting case is that of coupling constants $K_{j,j+1} = 1$
and quasiperiodic fields $h_j$. The ground state properties have recently been
analyzed exactly \cite{Sire}.
Here, we now want to consider non-periodic chains where $K_{j,j+1}$ {\em and} 
$h_j$ are modulated according to the two-letter substitution rule
$\varrho^{(k,\ell)}$ (\ref{3.1}) but taking only two possible values.
To be more precise, consider a particular
word $w_{n}$ ($w_{0}=a$, $w_{1}=b$, $w_{n+1}=\varrho^{(k,\ell)}(w_{n})$)
with length $N_{n}=f^{(k,\ell)}_n+kf^{(k,\ell)}_{n-1}$,
where the $f^{(k,\ell)}_{n}$ are defined (generalizing Eq.~(\ref{4.16})) by
\begin{equation}
f^{(k,\ell)}_{n+1}\;\; = \;\; \ell f^{(k,\ell)}_{n}\: +\:
k f^{(k,\ell)}_{n-1}
\hspace{1cm} f_{0}^{(k,\ell)} = 0, \hspace{5mm} f_{1}^{(k,\ell)} = 1 \; .
\label{6.5}
\end{equation}
To $w_{n}$ we associate an Ising spin chain with $N_{n} = N_{n}^{(k,\ell)}$
sites in the following way: We choose 
$K_{j,j+1} = K^{(i)}$ and $h_{j} = h^{(i)}$ ($i=0,1$) where
$i=0$ if the $j^{\mbox{th}}$ letter in the word $w_{n}$ is
an $a$ and $i=1$ otherwise. Hence the chains we consider are characterized
by four real parameters $K^{(i)}$ and $h^{(i)}$, $i=0,1$, and the two
numbers
$k, \ell \in\ZZ$ which determine the substitution rule.
Denoting the transfer matrix of the chain associated with the word
$w_{n}$ by ${\bm {\cal T}}_n$, one has the following recursive
equation for the transfer matrices
\begin{equation}
{\bm {\cal T}}_{n+1}\;\; =\;\; {\bm {\cal T}}_{n}^{\,\ell}\: 
{\bm {\cal T}}_{n-1}^{\, k}
\label{6.6}
\end{equation}
with initial conditions
\begin{equation}
{\bm {\cal T}}_{i}\;\; =\;\; \left( 
\begin{array}{ll} 
\exp\left( K^{(i)} + h^{(i)}\right) & \exp\left( -K^{(i)}\right) \\
\exp\left( -K^{(i)}\right) & \exp\left( K^{(i)} - h^{(i)}\right) 
\end{array}
\right) \hspace{1cm} i=0,1
\label{6.7}
\end{equation}
corresponding to the words $w_{0}=a$ and $w_{1}=b$.

Now, we are almost in the position to use the trace map
(\ref{3.2}) to obtain a recursive relation for the partition function
and hence for the free energy of our chains. In order to do so,
we define unimodular matrices
$\tilde{\bm {\cal T}}_{n} = \frac{1}{d_{n}} {\bm {\cal T}}_{n}$
where $d_{n} = \sqrt{\det {\bm {\cal T}}_{n}}$ and use the trace map
(\ref{3.2}) for 
\begin{equation}
x_{n} \;\; = \;\; \frac{1}{2}\tr\left( \tilde{\bm {\cal T}}_{n} \right)
\;\; = \;\; \frac{1}{2d_{n}} Z_{n}
\label{6.8}
\end{equation}
which yields a recursion relation for the partition function 
$Z_{n}=\tr ({\bm {\cal T}}_{n})$ of the chain associated with the word $w_{n}$,
compare \cite{Luck4} for an equivalent formulation with a nested iteration.
The determinants are simply given by (see Eq.~(\ref{6.5}))
\begin{eqnarray}
d_{n+1} & = & d_{n}d_{n-1} \nonumber \\
& = & (d_{0})^{kf^{(k,\ell)}_{n-1}} (d_{1})^{f^{(k,\ell)}_{n}} 
\label{6.9} 
\end{eqnarray} 
with $d_{i}=\sqrt{2\sinh (2K^{(i)})}$, $i=0,1$. 
In particular, all $d_{n}$ are non-zero if $d_{0}$ and $d_{1}$ are
different from zero, i.e. if both coupling constants $K^{(i)}$,
$i=0,1$, do not vanish. Thus the trick we used to be able to apply
the trace map (\ref{3.2}) to the transfer matrices of the Ising chain was
to 
split the recursion relation (\ref{6.6}) for the transfer matrices into
two parts: one for the trace of the transfer matrices and a rather trivial 
one for their determinant.

Explicitly, the trace map (\ref{3.2}) becomes
\begin{equation}
\left( \begin{array}{c}
 x_{n} \\
 y_{n} \\
 z_{n}
\end{array} \right)
\hspace{4mm}
\longrightarrow
\hspace{4mm}
\left( \begin{array}{c}
 x_{n+1} \\
 y_{n+1} \\
 z_{n+1} \\
\end{array} \right)
\hspace{2mm}
=
\hspace{2mm}
\left( \begin{array}{c}
 y_{n} \\
 g(x_{n},y_{n},z_{n}) \\
 h(x_{n},y_{n},z_{n}) \\
\end{array} \right),
\label{6.10}
\end{equation}
compare Eq.~(\ref{3.2}),
with starting values
\begin{eqnarray}
 x_{0} & = & \frac{\tr\left({\bm {\cal T}}_{0}\right)}{2d_{0}} 
\hspace{4mm} =\hspace{4mm}  
\frac{\exp K^{(0)} \cosh h^{(0)}}
     {\left(2\sinh (2K^{(0)})\right)^{1/2}} \nonumber \\
 y_{0} & = & \frac{\tr\left({\bm {\cal T}}_{1}\right)}{2d_{1}}
\hspace{4mm} =\hspace{4mm}  
\frac{\exp K^{(1)} \cosh h^{(1)}}
     {\left(2\sinh (2K^{(1)})\right)^{1/2}} \label{6.11} \\
 z_{0} & = & 
  \frac{\tr\left({\bm {\cal T}}_{0}{\bm {\cal T}}_{1}\right)}{2d_{0}d_{1}}
  \nonumber \\
       & = & 
\frac{\exp (K^{(0)}+K^{(1)}) \cosh (h^{(0)}+h^{(1)}) + 
      \exp(-(K^{(0)}+K^{(1)}))} 
{2\left(\sinh (2K^{(0)})\sinh (2K^{(1)})\right)^{1/2}}. \nonumber
\end{eqnarray}
Iterating the trace map thus corresponds to performing the
thermodynamic limit $n\rightarrow\infty$.

In this way the trace map (\ref{3.2}) provides an efficient 
way to compute the free energy in the
infinite size limit numerically. 
Unfortunately, an analytic solution to the recursion
relation is only known for the case that the starting point
lies on the invariant variety 
${\cal M}_0 = \{ (x,y,z) \in \CC^3 \which I(x,y,z) = 0 \}$ which turns out
to be equivalent to the requirement that the transfer matrices 
${\bm {\cal T}}_{0}$ and ${\bm {\cal T}}_{1}$ commute.
For this case, the solution is trivial since one
only counts the relative frequencies of the 
two different couplings in the chain.

It should be added that recursion relations for
the magnetization (one-point function) can be derived by similar means.
So far, we have not been able to extend this procedure 
to higher correlation functions.
A renormalization group analysis 
(see \cite{NelsonFisher}) of the field-free case
($h^{(0)}=h^{(1)}=0$)
gives the substitution matrix as linearization
at the zero-temperature fixed point and hence the thermal exponent $\alpha=1$
as one would expect. Further work along these directions is in progress.

\section{Concluding remarks}
\setcounter{equation}{0}
\label{sec7}

Having described structure and applications of trace maps of two-letter
substitution rules, we will briefly address further developments. It seems
that the mathematical aspects are understood fairly well.
Nevertheless, the trace maps as dynamical systems on their own right show
interesting features that deserve further exploration: orbit structure,
nontrivial symmetries, reversibility, period-doubling, and pseudo Anosov
structure, compare \cite{Roberts} for some results.

The physical applications could still profit from simply soaking up known
exact results. A prominent example is the gap labeling where the combination
of algebraic methods, generalized Bloch theory and pertubation theory
improves the understanding of spectra and wave functions. For the latter,
the extension to matrix maps should also be studied in more detail.
A similar remark holds true of the other applications. Kicked two-level
systems and spin systems are not yet brought to the understanding possible.

Another, rather restrictive point is the treatment of two-letter rules:
more than that is desirable. Several results are known for $n$-letter
substitution rules, compare \cite{Belli1,Graham,Peyriere2,Luck3,Franz},
but the generality of the findings is lacking. Many simple properties do not
extend, e.g., trace maps are higher dimensional and investigation of invariants
is much more complicated. Only the gap labeling theorem still applies in full
generality, wherefore further work in this direction will be needed.

\section*{Acknowledgements}
\label{sec8}

The authors are grateful to A.\ van Elst, P.\ Kramer, J.\ A.\ G.\ Roberts,
C.\ Sire, and F.\ Wijnands for discussions and helpful comments.
Financial support from Deutsche Forschungsgemeinschaft and Alfried Krupp
von Bohlen und Halbach Stiftung is thankfully acknowledged. M.\ B.\ would
like to thank the Department of Mathematics at the University of Melbourne
for hospitality where part of this work was done.
It is a pleasure to dedicate this article to Peter Kramer on the
occasion of his 60th birthday and to thank him this way for the
support over the years.
\section*{Appendix}
\renewcommand{\theequation}{A.\arabic{equation}}
\setcounter{equation}{0}
\label{app}

If $F_{\varrho}$ is the trace map of a two-letter substitution rule 
$\varrho \in \mbox{Hom}(F_2)$, then there is a uniquely defined transformation
polynomial $P_{\varrho} \in \ZZ[x,y,z]$ such that the Fricke character
$I(x,y,z) = x^2 + y^2 + z^2 - 2xyz - 1$ shows the transformation law
\begin{equation}
\label{app.1}
I(F_{\varrho}(x,y,z)) = P_{\varrho}(x,y,z) \cdot I(x,y,z)
\end{equation}
Here, the standard basis is used, i.e.,
$x = \frac{1}{2}\tr(\AAA), y = \frac{1}{2}\tr(\BB),
z = \frac{1}{2}\tr(\AAA\BB)$ with $\AAA, \BB \in \mbox{Sl}(2,\CC)$.
Instead of the complex numbers, one can also use
any other algebraically complete field.

For the proof of this remarkable equation, the variety
${\cal M}_0 = \{(x,y,z) \in \CC^3 \which I(x,y,z) = 0 \}$ plays an important
role. We first observe that
\begin{equation}
\label{app.2}
\tr(\KK(\AAA,\BB)-\ID) = 4 \cdot I(x,y,z)
\end{equation}
for unimodular 2x2-matrices, where $\KK(\AAA,\BB) = \AAA\BB\AAA^{-1}\BB^{-1}$
is the group commutator. This follows straightforwardly from the
Cayley-Hamilton theorem. We observe next
that any triple $(x,y,z) \in {\cal M}_0$ can be seen as traces of diagonal
matrices since we can simply choose $\AAA = \mbox{diag}(\alpha,\alpha^{-1}),
\hspace{2mm} \BB = \mbox{diag}(\beta^{\varepsilon},\beta^{-\varepsilon})$,
with $\alpha = x + \sqrt{x^2 + 1}, \hspace{2mm} \beta = y + \sqrt{y^2 + 1}$,
and $\varepsilon = \pm1$. Then, we find from the condition $I(x,y,z) = 0$
(a quadratic equation in $z$), the relation
$2z = \alpha\beta^{\varepsilon} + 1/(\alpha\beta^{\varepsilon})$, i.e.,
$z = \frac{1}{2}\tr(\AAA\BB)$. The two choices of $\varepsilon$ cover the
two possible solutions of the quadratic equation.

Observing that $[\AAA,\BB] = 0$ implies $\KK(\AAA,\BB) = \ID$, we see from
Eq.~(\ref{app.2}) that $I(x,y,z) = 0$ implies $I(F_{\varrho}(x,y,z)) = 0$. On the
other hand, we know that
\begin{equation}
\label{app.3}
P_{\varrho}(x,y,z) = \frac{I(F_{\varrho}(x,y,z))}{I(x,y,z)}
= \frac{\tr(\KK(\varrho(\AAA),\varrho(\BB))-\ID)}
{\tr(\KK(\AAA,\BB)-\ID)} .
\end{equation}
Thus, $P_{\varrho}$ is the quotient of two polynomials with integral
coefficients.
The denominator, $I(x,y,z)$, is irreducible in $\ZZ[x,y,z]$, and the numerator
polynomial vanishes at least on the set ${\cal M}_0$. From this, we can
conclude that the denominator divides the numerator. Since this can been seen
in an elementary way, we will briefly give the argument.

Consider a polynomial $Q \in \CC[x,y,z]$ that vanishes at least on
${\cal M}_0$. If $Q$ is of less than second order in $x, y$, or $z$, we can
only have $Q \equiv 0$. To show this, assume, without loss of generality,
$Q$ linear in $z$, i.e., $Q = \xi(x,y) + \eta(x,y) \cdot z$ with
polynomials $\xi$ and $\eta$.
For arbitrary $x, y \in \CC$ we can find $z_{\pm} = z_{\pm}(x,y)$ so that
$(x,y,z_+) \in {\cal M}_0$ and $(x,y,z_-) \in {\cal M}_0$. Consequently, $Q$
must also vanish on these points, which gives
\begin{equation}
\label{app.4}
z_+ \cdot \eta(x,y) = z_- \cdot \eta(x,y).
\end{equation}
But we can choose $(x,y)$ such that $z_+(x,y) \not= z_-(x,y)$ wherefore
$\eta(x,y) = 0$.
Furthermore, there is a whole $\varepsilon$-disc in $\CC^2$ around $(x,y)$
with $z_+(x,y) \not= z_-(x,y)$. On this disc, $\eta$ vanishes
due to Eq.~(\ref{app.4}). Thus
$\eta \equiv 0$ on $\CC^2$ and, from $Q(x,y,z_+) = 0$, also 
$\xi \equiv 0$. Hence, $Q \equiv 0$ under the conditions stated.

Let us now consider $Q = I(F_{\varrho}(x,y,z))$, which vanishes at least on
${\cal M}_0$. If $Q$ is only linear (or less) in $z$, say, the above
argument applies and we have $Q \equiv 0$ --- giving $P_{\varrho} \equiv 0$
as well. But, if $Q$ is at least quadratic in $z$, we can apply
Euler's division algorithm because the ring $\ZZ[x,y,z]$ is factorial
\cite{Lang} and the denominator starts with $z^2$. We find
\begin{equation}
\label{a5}
Q = P_{\varrho} \cdot I + r
\end{equation}
with $r$ a polynomial of degree $\leq 1$ in $z$. But $r$ obviously vanishes
at least
on ${\cal M}_0$, wherefore we must have $r \equiv 0$ according to the above
given argument.

We have thus rigorously and completely established that $I(x,y,z)$ devides
$I(F_{\varrho}(x,y,z))$. Now, it remains to be stated that $\ZZ[x,y,z]$ is
factorial wherefore $P_{\varrho}$ is again a polynomial with integral
coefficients. This proves Eq.~(\ref{app.1}).

\newpage
\section*{Figures}

\vspace*{\fill}
\begin{figure}[h]
\centerline{\epsfxsize=\textwidth\epsfbox{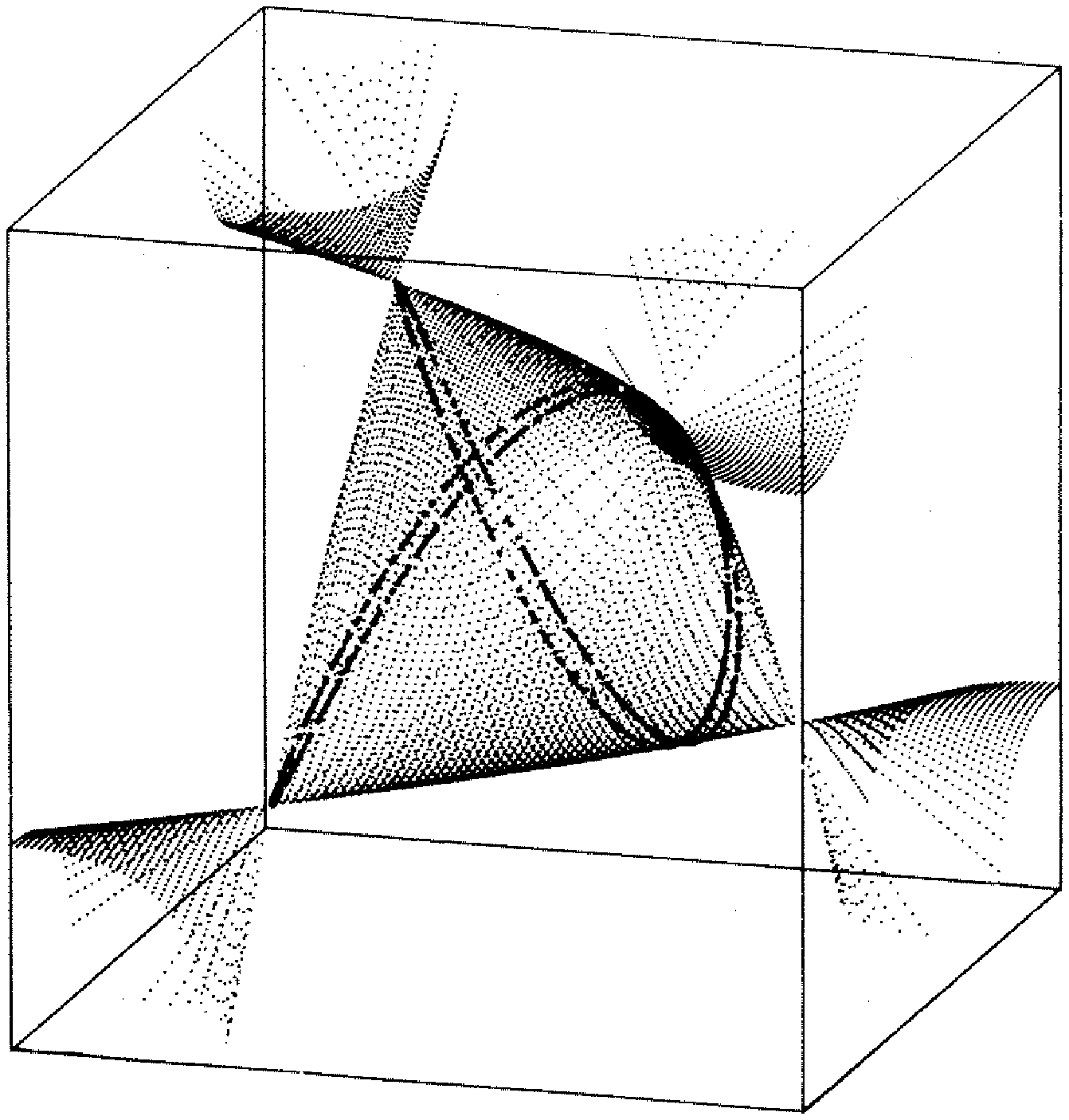}}
\caption{The invariant sheet $S = \{ I(x,y,z) = 0 \}$ and an 
orbit of the trace map ($k = 2, \ell = 1$)
lying also on the second invariant $H(x,y,z)$ which foliates $S$.}
\end{figure}\vspace*{\fill}\clearpage

\vspace*{\fill}
\begin{figure}[h]
\centerline{\epsfxsize=\textwidth\epsfbox{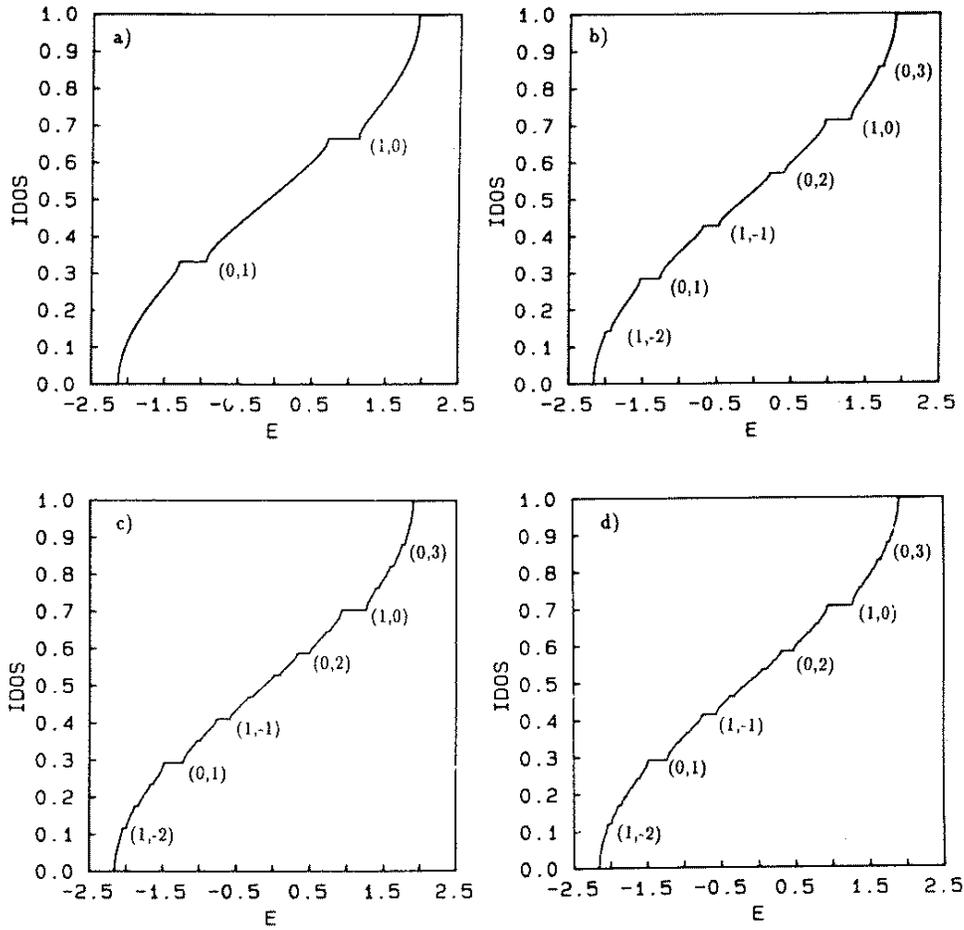}}
\caption{The approximants of the octonacci chain, 
$k = 1, \ell = 2$, with length
a) 3, b) 7, c) 17, d) 41 and the gap labels, $(\mu,\nu)$, 
starting with the 3- and the 7-approximant.}
\end{figure}\vspace*{\fill}

\end{document}